\documentclass[aps,prl,twocolumn,showpacs,superscriptaddress,groupedaddress]{revtex4}
\usepackage{graphicx}  
\usepackage{dcolumn}   
\usepackage{bm}        
\usepackage{amssymb}   
\usepackage{epstopdf}
\usepackage{amsmath}
\usepackage{nicefrac}
\usepackage{tikz}
\usepackage{enumitem}
\usepackage{comment}
\usepackage{array,multirow}
\usepackage[normalem]{ulem}
\usepackage{setspace}

\usepackage{hyperref}
\usepackage{soul}

\begin{document}

\title{
	Universal separable structure 
        of the optical potential 
 }
\author{H.~F.~Arellano}
\affiliation{Department of Physics-FCFM, University of Chile, Av. Blanco Encalada 2008, RM 8370449 Santiago, Chile}
\affiliation{CEA,DAM,DIF F-91297 Arpajon, France}
\author{G.~Blanchon}
\affiliation{CEA,DAM,DIF F-91297 Arpajon, France}
\affiliation{Universit\'e Paris-Saclay, CEA, LMCE, 91680 Bruyères-le-Ch\^atel, France}

\date{\today}

\begin{abstract}
  Based on a momentum-space \textit{in-medium} folding model,
  we disclose
  the universal separability of the optical potential,
  revealing its radial and nonlocality features
  at beam energies in the range 40–400 MeV and target mass numbers
  in the range $40\!\leq A\!\leq 208$.
  From this microscopic study
  we find that the nonlocality form factor is inherently complex
  and of hydrogenic nature, affecting both central and
  spin-orbit components of the potential.
  A striking outcome from this study is the consistent appearance
  of a nodal point in the imaginary radial form factor,
  notably suppressing surface absorption peaks,
  in evident contrast with Woods-Saxon's assumption of an absorptive peak
  at the nuclear surface.
  Our analysis reveals that the complex radial form factor can
  effectively be represented as convolutions of uniform spherical
  distribution with a Gaussian form factor and a Yukawa term.
  These 
	robust microscopically-driven
	findings offer new
  ways for investigating nuclear reactions
  beyond the restricting Woods-Saxon and Perey-Buck assumptions.
\end{abstract}

\maketitle

\section{Introduction}
Supernovae, nuclear spallation, stockpile stewardship, 
neutron star mergers, nuclear reactors, among many others, 
constitute scenarios where
multiple nuclear reactions involving stable and exotic nuclei take place.
In this context the optical potential emerges as a primary
tool to quantify scattering and reaction processes~\cite{Hebborn2023}.
Along this line,
phenomenological approaches inspired by Woods-Saxon
form factors~\cite{Woods1954},
supplemented with Perey-Buck nonlocal form 
factor~\cite{Perey1962},
have been developed with reasonable 
success~\cite{Perey1962,Koning2003,Morillon2007,Tian2015}.
Despite these advances, there is no theoretical justification
for the prescribed form factors associated to flux absorption nor
the nonlocality, a situation which may hinder
global models from further improvements.

From a fundamental stand point,
the challenge is to describe nucleon-nucleus
(\textit{NA}) scattering starting from the basic internucleon
interaction.
A final and ambitious goal would be to be able to describe 
scattering processes over a broad energy range 
--from a few keV up to hundreds of MeV-- 
and considering targets over the whole nuclear chart.
To this end we rely on \textit{ab initio} 
approaches~\cite{Hagen2012,Rotureau2017,Idini2019,Rotureau2020}
and \textit{in-medium} microscopic models~\cite{Arellano1995,Amos2000}.
Despite the significant advances reported in the context of 
\textit{ab-initio} approaches, current applications are limited
to few light targets and low energies, reaching qualitative
agreement with the data.
Microscopic models, on the other hand, constitute a reasonable
and flexible framework to bridge the basic nucleon-nucleon 
(\textit{NN}) interaction with the optical
potential by means of density-dependent \textit{NN} effective interactions.

Following a previous work, where a bell-shape nonlocality in the optical
potential was first disclosed~\cite{Arellano2022}, 
we further scrutinize its momentum-space
structure to extract its radial as well as nonlocal form factors.
The study we report here is based on the Argonne $v_{18}$ 
\textit{NN} potential (AV18)~\cite{Wiringa1995},
suited for \textit{NA}
scattering at the energies of this study, provided minimal relativity
is accounted for~\cite{Arellano2002}.
In this case the \textit{in medium} effective interaction is taken
from the genuine fully off-shell $g$ matrix solutions of the
Brueckner-Hartree-Fock integral equation. 
Emphasis is given to retain the genuine nonlocality of the 
optical potential at all stages of its calculation. 

This work is organized as follows.
In Sec. I we introduce the general framework of the the 
microscopic model, followed by Sec. III where we present
and discuss the
separability of the optical potential from actual folding
calculations. 
In Sec. IV we present final remarks.

\section{Framework}
The nonrelativistic optical-model potential in momentum space 
for nucleon elastic scattering off spin-zero nucleus 
can be expressed as~\cite{Ray1992}
\begin{equation}
  \label{ukk}
  \tilde {\cal U}({\bm k'},{\bm k};E) =
  \tilde {\cal U}_{c}({\bm k'},{\bm k};E) +
  i{\bm\sigma}\!\cdot\hat{\bm n}\,\,
  \tilde {\cal U}_{so}({\bm k'},{\bm k};E)\;,
\end{equation}
with $E$ the center-of-mass (c.m.) kinetic energy,
$\textstyle{\small\frac12}{\bm\sigma}$ the spin of the projectile and
$\hat{\bm n}$ 
a unit vector perpendicular to the scattering plane given by
$\hat{\bm n}\!=\!{\bm k'\!\times\!\bm k}/{|\bm k'\times\bm k|}$.
Here $\tilde {\cal U}_{c}$ and $\tilde {\cal U}_{so}$ 
denote the central and spin-orbit components of the potential
expressed in terms of the relative momenta ${\bm k}$ and ${\bm k}'$.
To calculate the potential we follow 
Refs. \cite{Arellano1995,Arellano2002,Aguayo2008},
based on an infinite nuclear-matter model to represent the 
\textit{in-medium} \textit{NN} effective interaction starting 
from realistic \textit{NN} bare potentials~\cite{Arellano2015}.
The use of the 
 Slater approximation~\cite{Arellano1990b,Arellano1995}
for the one-body 
mixed density yields the simplified form
\begin{align}
  \label{folding}
  \tilde {\cal U}({\bm k'},{\bm k};E) =&
  4\pi\!\!\sum_{\alpha=p,n}
  \int_0^{\infty}\!\!\! z^2dz\,\rho_\alpha(z) j_0(qz) 
  \nonumber \\
	\times\int\! d{\bm P}\, n_z(P) &
	\left\langle \textstyle{\frac12({\bm k}'\!-\!{\bm P})}|
    g_{\bar K}^{p\alpha} (E\!+\!\bar\epsilon)
 |\textstyle{\frac12({\bm k}\!-\!{\bm P})}\right\rangle_{_{\!\cal A}}\;,
\end{align}
with ${\cal A}$ denoting antisymmetrization
and $\rho_\alpha(z)$ 
the density of nuclear species $\alpha$ at coordinate $z$.
Additionally, the $g$ matrix is
evaluated at the isoscalar density $\rho(z)$,
coupling the projectile $p$ with target nucleon $\alpha$.
In the above $n_z(P)$ sets bounds for the off-shell 
contributions in the $g$ matrix at coordinate $z$.

Calculations of optical potentials as in Eq.~\eqref{folding}
rely on two main inputs:
proton and neutron ground-state densities, which in this work 
are taken from Hartree-Fock calculations of Ref.~\cite{Negele1970};
and the bare \textit{NN} potential, which we have chosen
AV18 because of its ability account for
\textit{NN} scattering data up to 320 MeV.
This bare interaction is used to calculate \textit{NN} $g$ matrices 
fully off-shell over a mesh of Fermi momenta $k_F$.
We refer the reader to Ref.~\cite{Arellano2015} for further details.
The scattering observables for the resulting nonlocal
optical potentials in momentum space are calculated using the
{\small SWANLOP} package~\cite{Arellano2021}, 
based on Ref.~\cite{Arellano2019},
treating nonlocalities in the presence of the Coulomb interaction
without approximations.

\begin{center}
\begin{figure}[ht]
\includegraphics[width=0.9\linewidth] {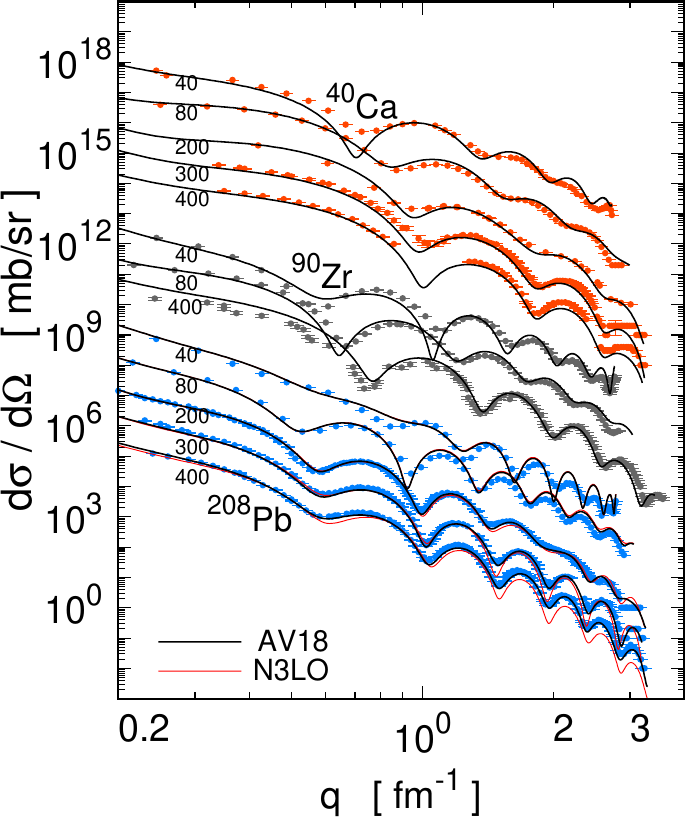}
\medskip
\caption{{\protect\small
\label{dsdw_i}
Differential cross section for proton-nucleus elastic scattering
as function of the momentum transfer based on 
    AV18 (black curves) and N3LO (red curves)
    bare potentials.
    The data are taken from
Refs.~\cite{Blumberg1966,Nadasen1981,Seifert1993,Hutcheon1988}.
        }
        }
\end{figure}
\end{center}

The consistency between the microscopic model and available data
is illustrated in Fig.~\ref{dsdw_i}, where we plot the 
measured~\cite{Blumberg1966,Nadasen1981,Seifert1993,Hutcheon1988}
and calculated~\cite{Arellano2021}
differential cross sections 
for proton-nucleus scattering
as functions of the momentum transfer.
The targets are $^{40}$Ca, $^{90}$Zr and $^{208}$Pb,
at the beam energies of 40, 80, 200, 300 and 400~MeV.
Black curves denote results based on AV18.
Being aware that chiral potentials are not suited for 
high energies, we still find instructive to illustrate 
its behavior at energies of this study.
Thus, we consider the chiral \textit{NN} potential up
to next-to-next-to-next-to-leading order (N3LO)~\cite{Entem2003}.
Results for $p+^{208}$Pb scattering
based on this chiral interaction are shown with red curves.
As observed, the folding model based on AV18
yields a reasonable description of the scattering data.
In the case of  40~MeV the calculated differential cross 
section appears slightly more diffractive than the data, 
leading us to set this as the lowest energy for our study.
In the case of the N3LO interaction for $p+^{208}$Pb
we observe a gradual departure of its results from the data,
an indication
of upper energy limit of about 290~MeV for this bare interaction.

\section{Separability}
The optical potential in Eq.~\eqref{folding}, 
expressed in terms of relative momenta $\bm k$ and $\bm k'$, 
can be reexpressed in terms of the momentum transfer ${\bm q}$  and
the mean momentum ${\bm K}$, namely
\begin{subequations}\label{kkqq}
\begin{align}
  {\bm K}&=\textstyle{\frac12} \, ( {\bm k} + {\bm k'}), \label{kk} \\
  {\bm q} &={\bm k} - {\bm k'} . \label{qq}
\end{align}
\end{subequations}
Hence, if we denote
$\tilde U({\bm K},{\bm q})\!=\!\tilde {\cal U}({\bm k'},{\bm k};E)$,
then
\begin{equation}
  \label{UKq}
{\tilde U}({\bm K},{\bm q})
	= {\tilde U}_{c}({\bm K},{\bm q}) +
  i {\bm\sigma}\cdot({\bm K}\times{\bm q})\, 
  {\tilde U}_{so}({\bm K},{\bm q})\;.
\end{equation}
Each component, 
${\tilde U}_{c}$ and ${\tilde U}_{so}$, can then be expanded
in terms of Legendre polynomials of 
 $w\!=\!\hat{\bm K}\cdot\hat{\bm q}$.
In this way we express
${\tilde U}_{c}({\bm K},{\bm q})\!=
\!\sum_{\substack{n=0}}^{\infty}{\tilde U}^{(c)}_{n}(K,q)P_n(w)$,
together with
$|{\bm K}\times{\bm q}| {\tilde U}_{so}({\bm K},{\bm q})\! =\!
 Kq\sum_{\substack{n=0}}^{\infty} 
 {\tilde U}^{(so)}_{n+1}(K,q) P^{1}_{n+1}(w)$.
In a recent study 
\cite{Blanchon2024} we found that the zeroth-order contribution 
in each expansion 
is sufficient to accurately describe the scattering observables 
of the original potential.
This remarkable property validates focusing solely on the $n\!=\!0$ terms,
justifying the notation 
$\tilde U_{c}({\bm K},{\bm q})\to \tilde U_{c}(K, q)$, and
$\tilde U_{so}({\bm K},{\bm q})\to \tilde U_{so}(K, q)$.

To isolate the nonlocality of $\tilde U_{c}(K,q)$ and 
$\tilde U_{so}(K,q)$, 
for now denoted as $\tilde U(K,q)$,
we define $J/(2\pi)^3\!=\!\tilde U(0,0)$, together with the ratios
\begin{subequations}
\begin{align}
  \label{VKq}
  \tilde V(K,q) &= \frac{\tilde U(K,q)}{\tilde U(K,0)}\,, \\
  \label{HK}
  \tilde H(K)&=\frac{\tilde U(K,0)}{\tilde U(0,0)}\;.
\end{align}
\end{subequations}
Note that $\tilde V(K,q)$ and $\tilde H(K)$
are complex and dimensionless, satisfying 
$\tilde V(K,0)\!=\!\tilde H(0)\!=\!1$.
Additionally, $J$ is also complex, 
representing the volume integral of the potential.
With these definitions $\tilde U(K,q)$ factorizes as
  \begin{equation}
  \label{separation}
  \tilde U(K,q) = \frac{J}{(2\pi)^3} \;\tilde V(K,q) \;\tilde H(K)\;.
  \end{equation}
 
In Ref.~\cite{Arellano2022} it was noted that $\tilde V(K,q)$ has
a weak dependence on $K$.
We have examined this dependency and found that by 
setting $\tilde V(K,q)\!=\!\tilde V(k_0,q)$, with $k_0$ 
the on-shell momentum in the c.m.  reference frame,
the factorization in Eq.~\eqref{separation} reproduces the scattering 
observables of the original potential.
By making this choice we ensure that the 
forward ($q\!=\!0$) on-shell matrix element of the potential is reproduced.
This result motivates the definition of the radial form
factor, $\tilde v(q)\!=\! \tilde V(k_0,q)$. 
The denomination
\textit{radial} 
refers to the fact that the conjugate coordinate of $q$ corresponds
to the mean radial distance between the projectile
and the center of the target, hereafter denoted as $r$.
The above considerations lead to 
  \begin{equation}
  \label{jvh}
  \tilde U(K,q) = \frac{J}{(2\pi)^3} \;\tilde v(q) \;\tilde H(K)\;,
  \end{equation}
to be referred as \textit{JvH} factorization.
We now examine each of its three terms. 

\subsection{The volume integral $J$}
\begin{figure}[ht]
\includegraphics[width=0.90\linewidth] {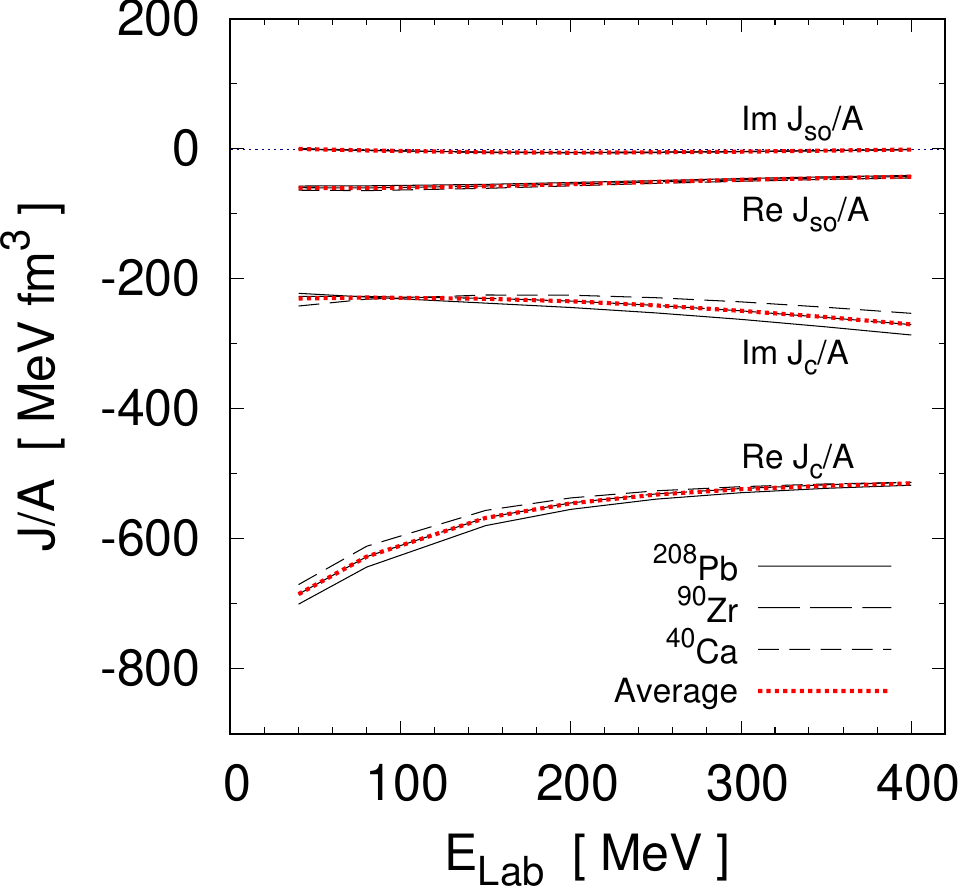}
\medskip
\caption{{\protect\small
\label{joa}
 Volume integrals per nucleon $J_c/A$ and $J_{so}/A$ as functions of the
	beam energy.
        }
        }
\end{figure}

In momentum space, the volume integral $J$ of the potential is obtained
from its matrix element at ${\bm k}\!=\!{\bm k}'\!=\!0$, 
that is to say $\tilde U(0,0)$.
In Fig.~\ref{joa} we plot the volume integrals of the potential 
over the number of target nucleons, $J/A$, as function of the 
beam energy.
In the following,
subscripts $c$ and $so$ denote central and spin-orbit components,
respectively.
Solid, long-dashed and dashed curves denote results for 
$^{208}$Pb, 
$^{90}$Zr and
$^{40}$Ca,
respectively. 
As observed, $\textrm{Re}\,[J_c/A]$ is the strongest term,
decreasing monotonically its magnitude by about 25\% over the range 
$40\!-\!400$~MeV.
The imaginary term, on the other hand, is much weaker but 
increasing its magnitude by $\sim\!17$\% over the same energy range.
The spin-orbit strengths $J_{so}/A$ are nearly real and constant,
amounting to less than 10\% of $\textrm{Re}\,[J_c/A]$.
We have verified that 
  the averaged $J_c/A$ and $J_{so}/A$, 
  considering all three targets and evaluated at each beam energy,
yield nearly the same scattering observables as the ones obtained 
from the original potential.
These averages are shown with dotted red curves. 

\subsection{The radial form factor $v$}
\begin{center}
\begin{figure}[ht]
  \includegraphics[width=0.90\linewidth] {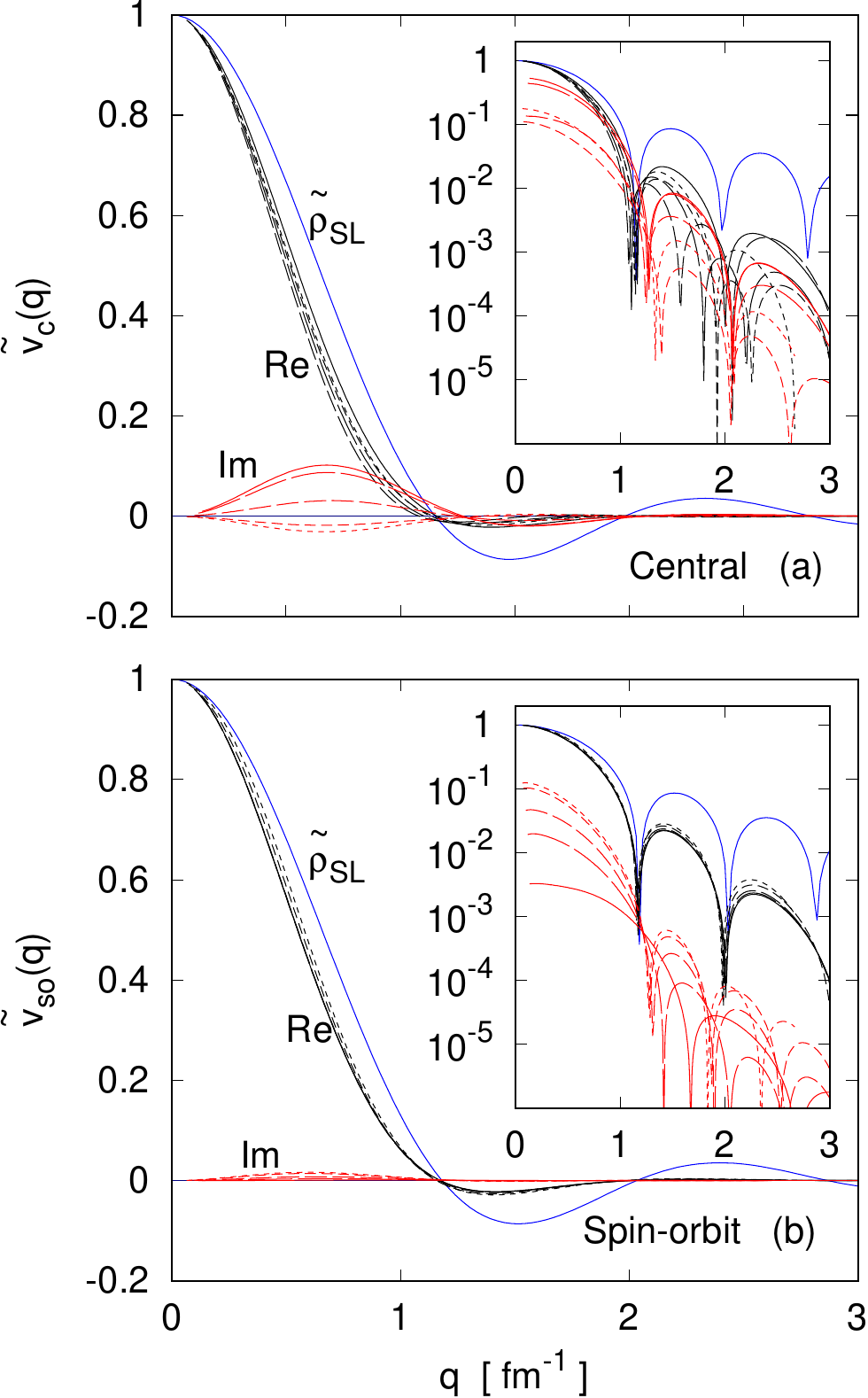}
\medskip
\caption{{\protect\small
\label{vq}
	Central (a) and spin-orbit (b) radial form factors as
	functions of the momentum transfer in the case of
	$p\!+\!^{40}\textrm{Ca}$ scattering.
        Results at 40, 80, 200, 300 and 400~MeV, are
shown with dotted, short-dashed, dashed, long-dashed and solid curves, 
respectively.
  Blue solid curves represent $\tilde\rho_{SL}(qR)$ (see text).
Red curves in the insets represent $|\textrm{Im\,}\tilde v(q)|/q^2$.
        }
        }
\end{figure}
\end{center}

We now examine 
the central $\tilde v_{c}(q)\!=\! \tilde V_{c}(k_0,q)$, 
and spin-orbit $\tilde v_{so}(q)\!=\! \tilde V_{so}(k_0,q)$, 
form factors.
In panels (a) and (b) of Fig.~\ref{vq} we show the central $\tilde v_c$ 
and spin-orbit $\tilde v_{so}$ radial potentials
as functions of $q$, respectively. 
These correspond to microscopic calculations for $p+^{40}$Ca
elastic scattering at 40, 80, 200, 300 and 400~MeV,
shown with dotted, short-dashed, dashed, long-dashed and solid curves, 
respectively.
The insets show logarithmic plots for 
$|\textrm{Re\;} \tilde v_c |$ and $|\textrm{Re\;}  \tilde v_{so} |$ 
(black curves), and the ratios
$|\textrm{Im\;} \tilde v_c/q^2 |$ and $|\textrm{Im\;}  \tilde v_{so}/q^2 |$ 
(red curves).
Blue solid curves correspond to 
$\tilde\rho_{SL}(qR)\!=\!3\,j_1(qR)/qR$,
the Slater density to be discussed later.

We observe similar diffractive patterns for the real components of
$\tilde v_c$ in panel (a), and $\tilde v_{so}$ in panel (b), 
with their first nodes at 
$q\!\approx\!1$ and 1.1~fm$^{-1}$, respectively.
In the case of their imaginary components we found that 
they behave as $\sim\!q^2$ near the origin, feature 
displayed in both insets, where red curves 
represent $\textrm{Im}\,[\tilde v(q)/q^2]$.
As observed, the diffractive pattern of this ratio is 
qualitatively similar to that of their real counterparts.
Apart from the exact location of the nodes, the
diffractive behavior of the Slater density $\rho_{SL}(qR)$ 
has an appealing resemblance with those shown with black and red curves
in the insets.
This observation becomes useful for the upcoming discussion.

\begin{center}
\begin{figure}[ht]
  \includegraphics[width=0.90\linewidth] {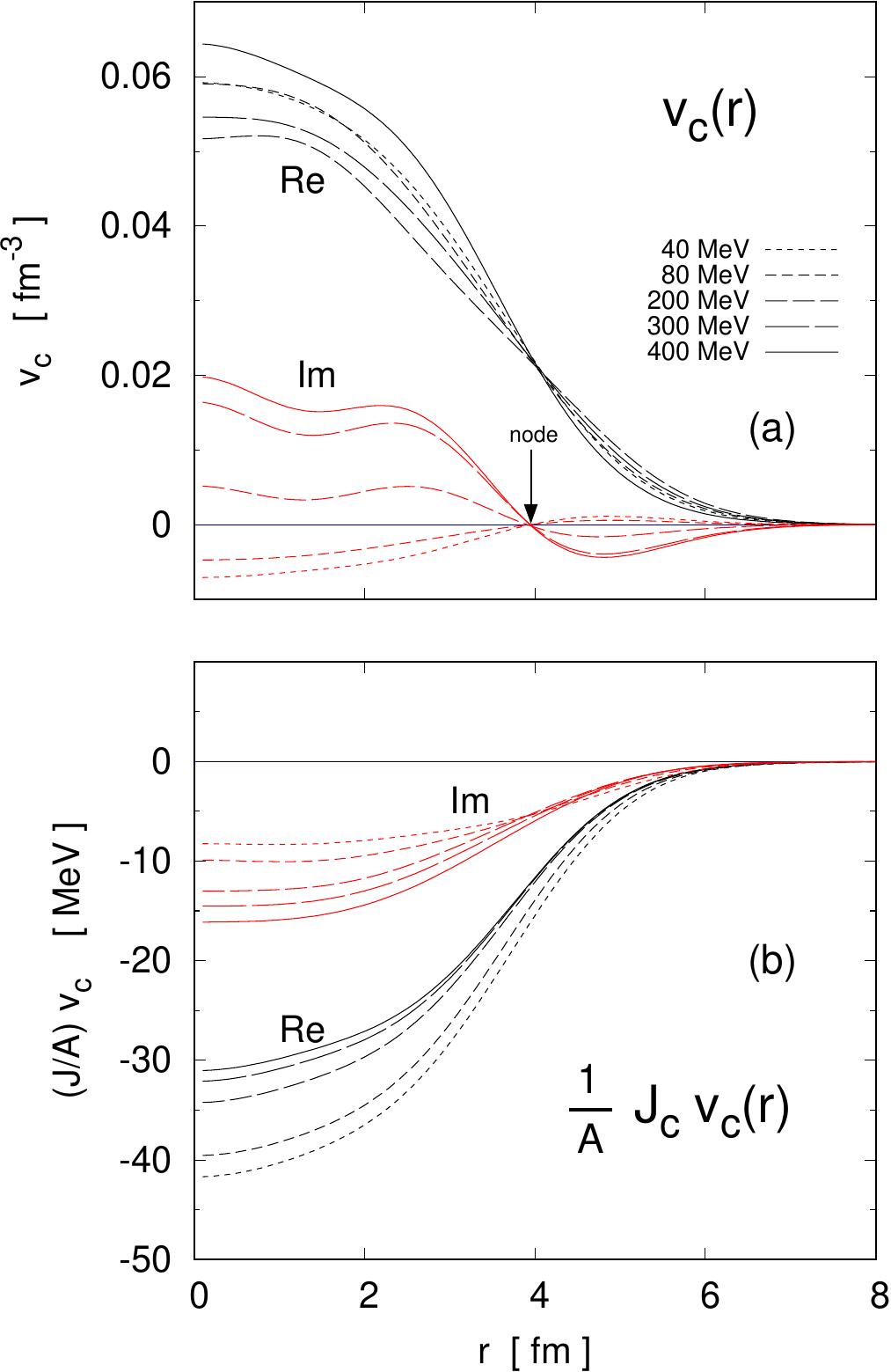}
\medskip
\caption{{\protect\small
\label{vr}
	Coordinate-space radial form factor $v_c(r)$ 
	(a)
	and 
        product $(J/A)v_c(r)$ 
	(b)
	as functions of $r$.
        Curve patterns follow the same convention as in Fig.~\ref{vq}.
        }
        }
\end{figure}
\end{center}

The radial form factors in coordinate space for the central
potential, $v_c(r)$, are displayed in panel (a) of Fig.~\ref{vr}.
These correspond to $p+^{40}$Ca scattering at the same energies
considered in Fig.~\ref{vq}, for which we adopt the same notation.
We note that all $\textrm{Re\;}[v_c(r)]$ resemble roughly a
two-parameter Fermi distribution (2pF).
Interestingly, the imaginary part of  $v_c(r)$
shows nodal points at nearly the same radius, $r\!\sim\!4$~fm.
These nodes are in correspondence with the change of sign
in the curvature
of $\textrm{Re\;}[v_c(r)]$, driven by its second derivative.
An interpretation of this feature emerges from the $q$ behavior
of $\textrm{Im\;}[\tilde v_c(q)/q^2]$ displayed with red curves
in the insets of Fig.~\ref{vq}, being qualitatively similar
to those for $\textrm{Re\;}[\tilde v_c(q)]$.
Thus, $q^2\,\textrm{Im\;}[\tilde v_c(q)/q^2]$ would account for nodes
near the surface of the 2pF distribution.
These peculiar features for $v_c$ are also observed for $v_{so}$.

The full potential in the \textit{JvH} form involves the product
of three complex terms.
The radial part of the optical potential is driven by $Jv$,
the product of the volume integral and the radial form factor,
being the natural counterpart of conventional Woods-Saxon potentials. 
Therefore, if $J\!=\!J_r+iJ_i$, and $v=v_r+iv_i$, then
the real part of $Jv$ involves a combination of both
real and imaginary parts of the radial form factor.
The same holds for $\textrm{Im\;}[Jv]$.
With this in mind, features of the radial form factor $v$ differ
from those of $Jv$, the radial potential.

In panel (b) of Fig.~\ref{vr} we plot the product
$(J/A)v_c$ as function of the radius.
As observed, the potential is attractive over the whole range, 
vanishing smoothly for increasing radius.
Even though the imaginary part of $(J/A)v_c$ also vanishes for 
increasing radius, at 40~MeV the absorption becomes stronger
above the surface ($r\gtrsim 4.2$~fm)
than that for higher energies.
Interestingly, $\textrm{Im\;}[Jv_c]$ shows no peak at the surface,
in contrast with the commonly used Woods-Saxon's prescription to 
model nuclear absorption phenomenologically.
These features for the absorption 
are also observed for the $^{90}$Zr and $^{208}$Pb targets.

\begin{center}
\begin{figure}[ht]
  \includegraphics[width=0.90\linewidth] {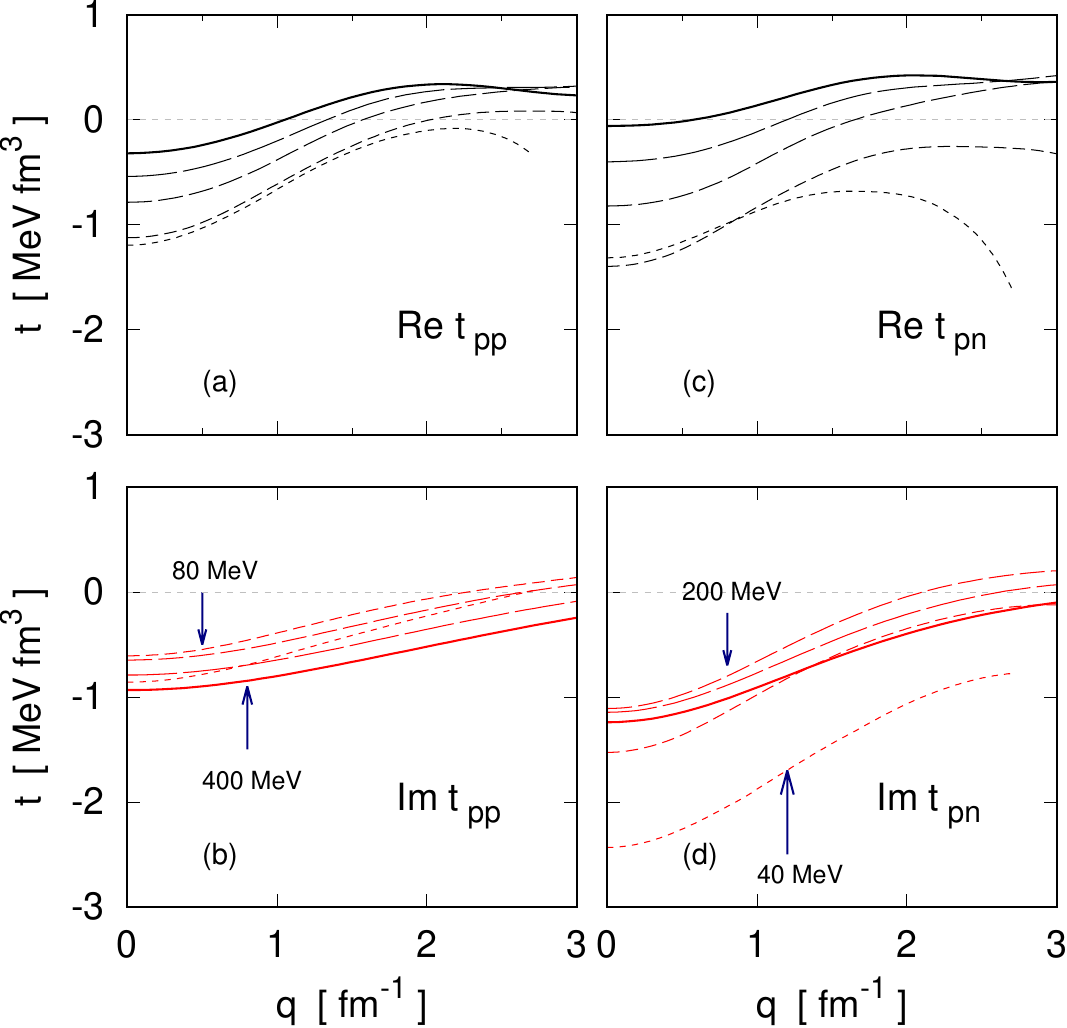}
\medskip
\caption{{\protect\small
\label{tmatrix}
 On-shell $t$ matrix for \textit{pp} and \textit{pn} channels
  as functions of the momentum transfer.
  Dotted, short-dashed, dashed, long-dashed and solid curves
  correspond to 40, 80, 200, 300 and 400~MeV, respectively.
  Black and red curves correspond to real and imaginary components,
  respectively.
        }
        }
\end{figure}
\end{center}

Going back to panel (a) in Fig.~\ref{vr} we notice that the
maxima of $\textrm{Re\,}v_c$, taking place at $r\!=\!0$, 
is not a monotonic function of the energy.
They decrease from 40~MeV up to 200~MeV, followed by a sudden increase
evidenced at 400~MeV.
This trend is consistent with $\tilde v_c(q)$ shown in panel (a) 
of Fig.~\ref{vq}, with the 400~MeV-case (solid curves) 
above all the other curves. 
We have investigated the origin of this trend and found that
it stems from the behavior of the effective interaction at different
energies. To illustrate the point, we have considered the
zero-density $g$ matrix in Eq.~\eqref{folding}, corresponding to
the free $t$ matrix.
In this case we examine the on-shell element
\begin{equation}
  \label{ttmatrix}
  t_{pp,pn}(q) =
	\left\langle \textstyle{\frac12{\bm k}'}|
    g_{\bar K}^{pp,pn} (E)
 |\textstyle{\frac12{\bm k}}\right\rangle_{_{\!\cal A}}\;,
\end{equation} 
where $\bm q\!=\!\bm k\!-\!\bm k'$, with $k=k'$, 
chosen on-shell in the \textit{NA} c.m.
In Fig.~\ref{tmatrix} we show the $t$ matrix as function of
the momentum transfer $q$ for the \textit{pp} channel 
[panels (a) and (b)] and \textit{pn} channel 
[panels (c) and (d)].
The upper and lower panels show results for the real and imaginary
components, respectively.
We observe that the real parts of $t$ follow a smooth
behavior as functions of $q$, with clear separation
among the curves as the energy increases.
Such is not the case for the imaginary components, where
the upper and lower curves follow no uniform order,
as indicated by the arrows in panels (b) and (d).

\begin{center}
\begin{figure}[ht]
  \includegraphics[width=0.90\linewidth] {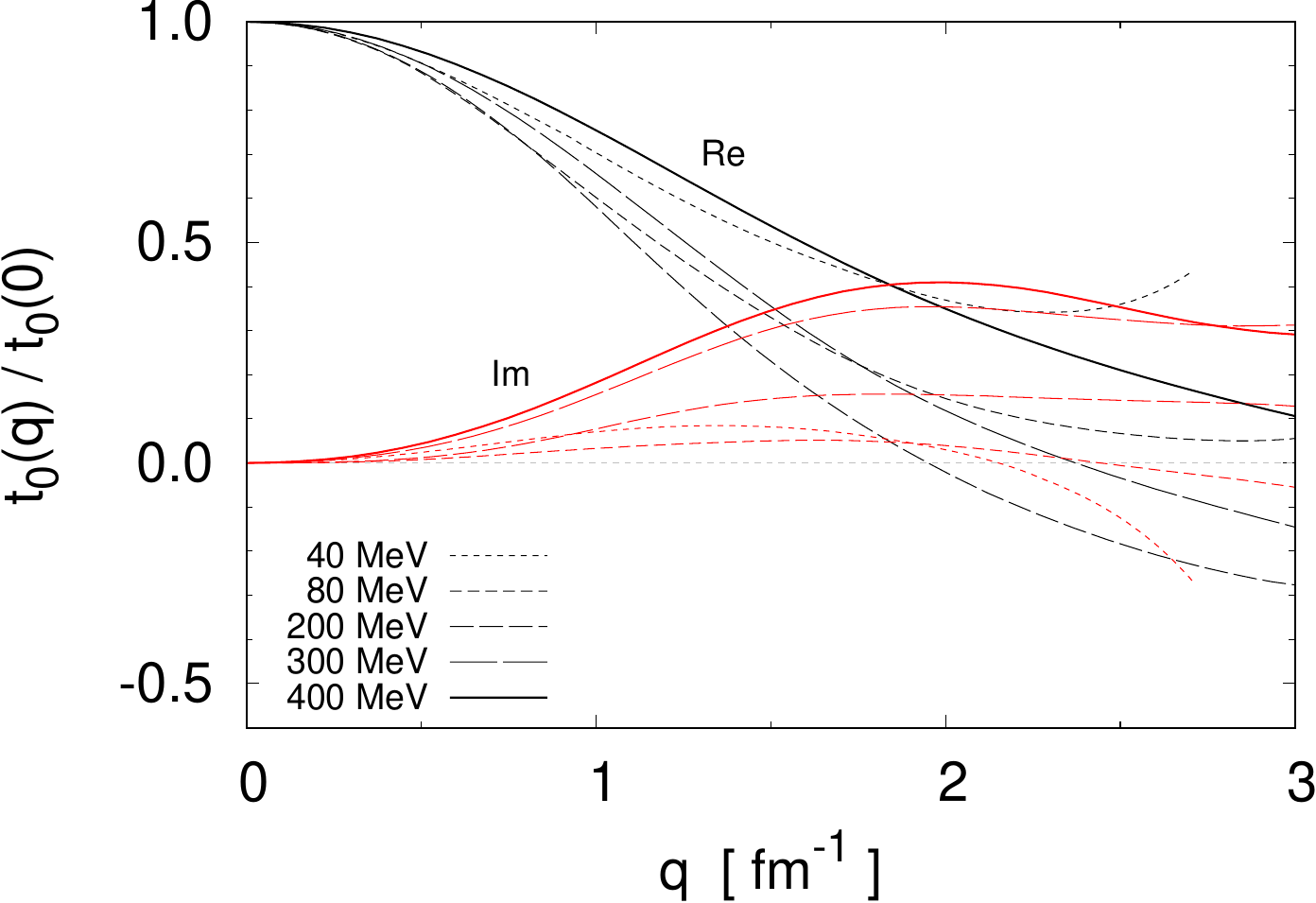}
\medskip
\caption{{\protect\small
\label{tmoby}
 The isoscalar ratio $t_{0}(q)/t_{0}(0)$ as function of the
  momentum transfer at the same energies as in Fig.~\ref{tmatrix}.
  Dotted, short-dashed, dashed, long-dashed and solid curves
  correspond to 40, 80, 200, 300 and 400~MeV, respectively.
  Black and red curves denote real an imaginary components, respectively.
        }
        }
\end{figure}
\end{center}

Having in mind the "$t\rho$" approximation for the optical
potential~\cite{Ray1992} together with Eq.~\eqref{VKq}
for $\tilde V(K,q)$, we observe that the radial form factor
$\tilde v_c(q)$ is closely related to the ratio $t(q)/t(q\!=\!0)$.
Thus, we have calculated this ratio for the isoscalar channel,
where $t_0\!=\!t_{pp}\!+\!t_{pn}$. 
In Fig.~\ref{tmoby} we plot $t_0(q)/t_0(0)$ at
the same energies considered in Fig.~\ref{tmatrix}.
It is interesting to note how the curve for 400~MeV stays above
the rest, in similar fashion as that for $\tilde v_c(q)$ in Fig.~\ref{vq}. 
This explains why $v_c(0)$ is maximum in the case of 400~MeV,
in addition to providing arguments for the non-monotonicity
of the maxima in $v_c(r)$. 

\subsection{The nonlocality form factor $H$}
\begin{center}
\begin{figure}[ht]
\includegraphics[width=0.90\linewidth] {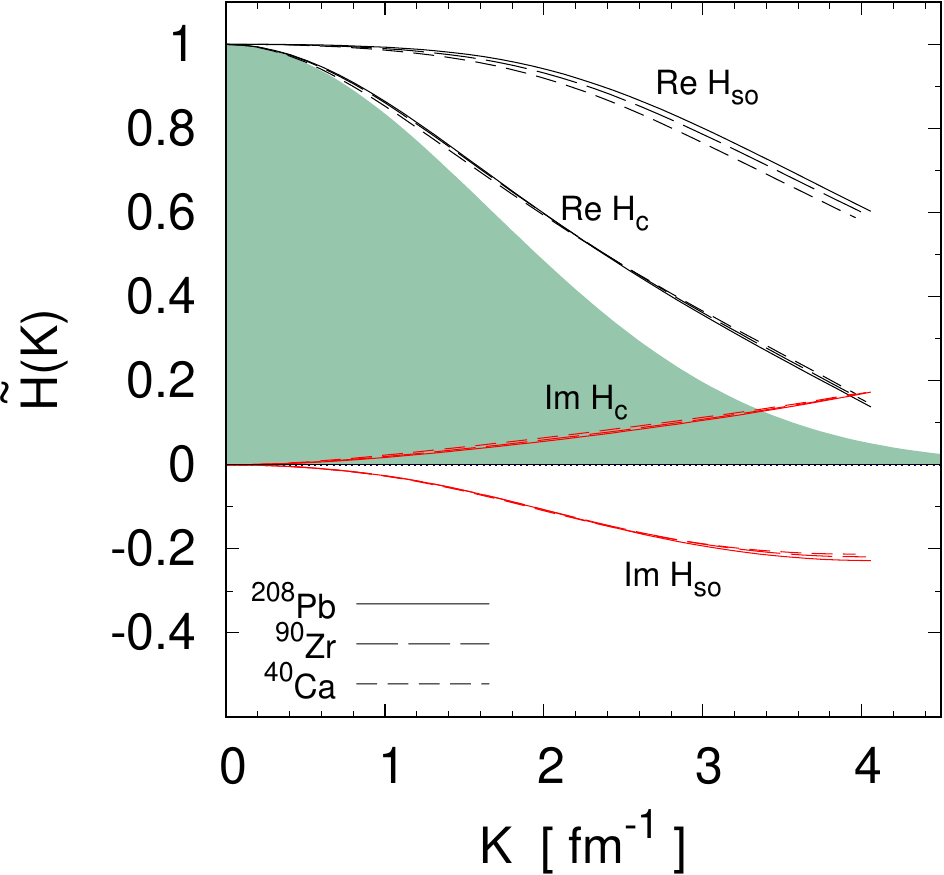}
\medskip
\caption{{\protect\small
\label{hk}
Bell-shape nonlocality form factor $\tilde H(K)$ for
\textit{NA} scattering at 200~MeV. 
The shaded area denote Perey-Buck nonlocality form factor
with $\beta\!=\!0.84$~fm.
        }
        }
\end{figure}
\end{center}

In Fig. \ref{hk} we plot the resulting
nonlocality form factors $\tilde H_c$ and $H_{so}$ as
functions of $K$, 
associated to proton-nucleus scattering at 200~MeV.
The targets are 
$^{208}$Pb, $^{90}$Zr and $^{40}$Ca,
following the same pattern convention as in Fig.~\ref{joa}.
The green-shaded profile corresponds to Perey-Buck's nonlocality
form factor given by
\begin{equation}
\tilde H_{PB}=\exp(-\beta^2K^2/4),
\end{equation}
with $\beta\!=\!0.85$~fm,
a typical value.
We observe that all targets show very similar nonlocal behavior,
featuring a weaker decay relative to the Gaussian nonlocality.

Aiming to obtain a simple and accurate representation for $\tilde H$,
we have found useful to represent it as
\begin{equation}
	\label{Hetalpha}
	\tilde H(K) = 
	\frac{1}{[1+\eta(K)]^2}
	\frac{1+i\alpha(K)}{1-i\alpha(K)}\;,
\end{equation}
with $\eta$ and $\alpha$ real functions 
obtained directly from $\tilde H$.

\begin{center}
\begin{figure}[ht]
  \includegraphics[width=0.90\linewidth] {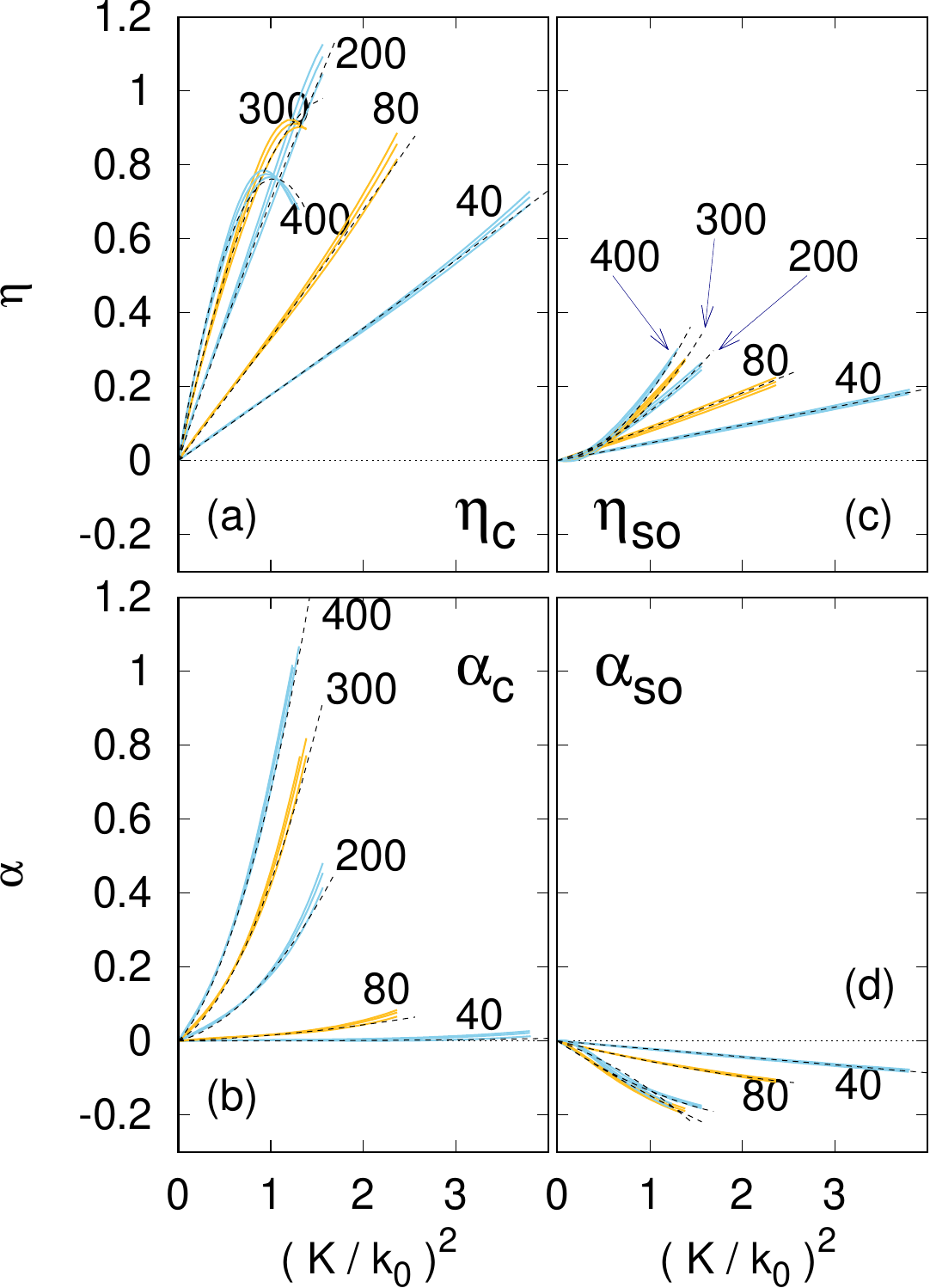}
\medskip
\caption{{\protect\small
\label{etalpha}
Nonlocality parameters $\eta$ and $\alpha$ as functions of $(K/k_0)^2$.
See text for explanation of curve patterns.
        }
        }
\end{figure}
\end{center}

In Fig. \ref{etalpha} we plot the resulting central and spin-orbit
$\eta$ and $\alpha$ as functions of $(K/k_0)^2$, with $k_0$
the relative momentum in the c.m. reference frame.
Numerical labels on each graph 
denote beam energy in units of MeV.
Each bunch of colored curves include results for each of the three targets.
We observe that, at a given energy, all three targets yield similar
behavior in their respective $\eta$ and $\alpha$.
Thus, 
to each of these sets we 
give an optimal quadratic form of the type
$(b_1x\!+\!b_2x^2)$, with $x\!=\!(K/k_0)^2$.
The resulting parametrizations are shown with dotted curves in 
Fig. \ref{etalpha},
reproducing the scattering observables of the original potential.

The coordinate-space nonlocality form factor is obtained from 
\begin{equation}
H(s)={\frac{1}{2\pi^2}}
\int_{0}^{\infty}K^2dK\,j_0(Ks)\tilde H(K)\,,
\end{equation}
  with $s\!=\!|\bm r'-\bm r|$, the difference between post and
  prior \textit{NA} relative coordinates in Perey-Buck's 
  notation~\cite{Perey1962}.
To gain some insight from Eq.~\eqref{Hetalpha} for $H$ we
consider the lowest order term in the expansions for $\eta$ and $\alpha$.
Hence, by expressing
$\eta\!\approx\! b^2K^2$, and $\alpha\!\approx\! a^2K^2$,
contour integration in the complex plane yields the closed form
\begin{equation}
\label{rspace}
	H(s)=
	\frac{
          \left ( 
        a^4+b^4
        +
        4i\frac{a^2 b^3}{s}
        \right )
          e^{-s/b}
        -
        4i\frac{a^2 b^3}{s} e^{-s\sqrt{i}/a }}
	      {8\pi b^3(b^2+ia^2)^2}\;.
\end{equation}
The real-argument exponential $e^{-s/b}$ accounts for nonlocality of 
hydrogenic type of range $b$, 
in contrast with the commonly used Gaussian form factor introduced 
by Perey and Buck.
The complex-argument exponential, on the other hand, 
expresses a strongly dumped oscillatory contribution of range $a\sqrt{2}$. 
We also note that the nonlocality $H$ is finite at the origin, 
and that for vanishing $a$ it becomes 
\begin{equation}
  H(s)=\frac{e^{-s/b}}{8\pi b^3}.
\end{equation}

\subsection{Global representation}
  In the preceding sections we have examined the leading features
  exhibited by each of the three terms participating in the
  \textit{JvH} structure of the optical potential.
  In this section we explore a possible representation
  of the radial form factors, $v_c$ and $v_{so}$, retaining
  the microscopic $J/A$ and nonlocality $\tilde H(K)$ expressed
  by Eq.~\eqref{Hetalpha}.
  The aim here is to reproduce as closely as possible
  the microscopic radial form factors $v$ and with that
  --through the $JvH$ separable structure of the potential--
  the scattering observables of the original folding potential.

To obtain a suitable representation for the microscopic
$\tilde v_c$ and $\tilde v_{so}$, we find useful to consider
the Slater density $\tilde\rho_{SL}(qR)$,
the three-dimensional Fourier transform of a uniform sphere of radius $R$.
The behavior of $\tilde\rho_{SL}$ is shown in Fig.~\ref{vq}
with blue curves, 
with $R$ calibrated to match the first zero of 
the spherical Bessel function $j_1$.
Although there is no full correspondence between the zeros of 
$\tilde v(q)$ and $\tilde\rho_{SL}(qR)$, we have found that by folding
additional form factors it is possible to reasonably reproduce
the scattering observables of the original microscopic potential.
Thus, we have explored the following convolutions:
\begin{subequations}\label{fits}
\begin{align}
  \textrm{Re}\,[\tilde v_{c,so}(q)] &= \tilde\rho_{SL}(qR_x)
           \,e^{-a_x^2q^2}\,\frac{1}{1+b_x^2q^2} ,
  \label{rev}
           \\
  \textrm{Im}\,[\tilde v_{c,so}(q)] &= c_y\, q^2 \tilde\rho_{SL}(qR_y)
           \, e^{-a_y^2q^2}\,\frac{1}{1+b_y^2q^2} .
   \label{imv}
\end{align}
\end{subequations}
This construction can be interpreted as if $\tilde v_{c,so}$ results
from the convolution of a Gaussian-smoothed uniform sphere with a
Gaussian-dressed one-meson-exchange interaction.
The product of the two Gaussian form factors, also Gaussian,
would then account for both the smoothing and the dressing.

\begin{center}
\begin{figure}[ht]
\includegraphics[width=0.90\linewidth] {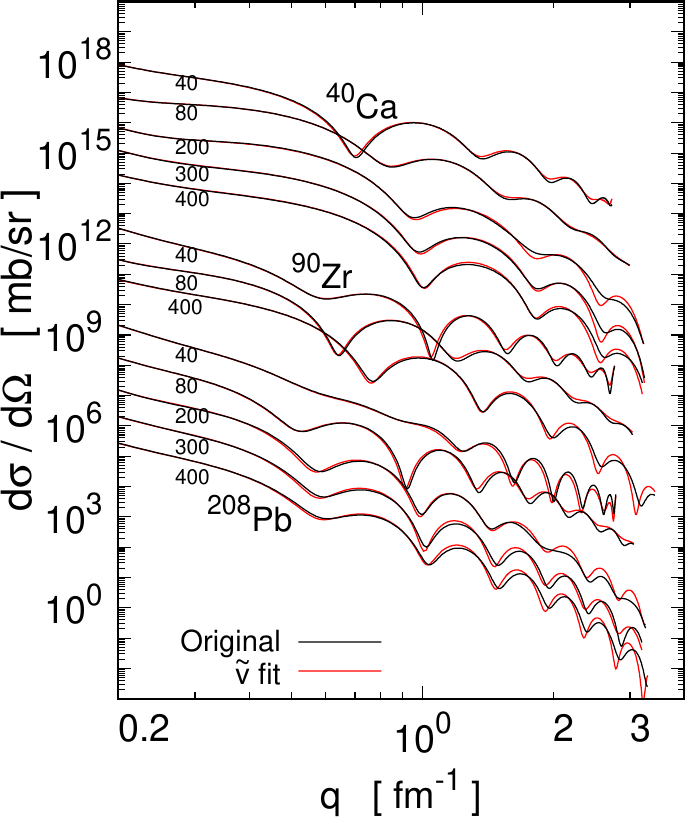}
\medskip
\caption{{\protect\small
\label{dsdw_ii}
Differential cross sections as functions of the momentum transfer
based on the microscopic folding model (black curves) and global 
	fit for $\tilde v(q)$ (red curves).
        }
        }
\end{figure}
\end{center}

We have performed a search of 
parameters $R_{x,y}$, $a_{x,y}$, $b_{x,y}$ and $c_y$, 
for the central and spin-orbit components, 
to globally reproduce the microscopic $\tilde v$ 
at the five beam energies and the three targets considered in this study.
The ability of these global convolutions
to reproduce scattering observables of the original microscopic 
potential is illustrated in Fig.~\ref{dsdw_ii},
showing with red curves the resulting $d\sigma/d\Omega$
and black curves the ones from the original potential.
Although red and black curves do not fully overlap, 
we note a close correspondence between them.
Improvements can be envisaged if separate analyses are made
to the coupling to target protons and neutrons, in addition 
to alternative form factors and search strategies.

\section{Concluding remarks}
The universal separability of the optical potential 
is grounded on two elements.
The first one is its general form, expressed as 
$\tilde{\cal U}(\bm k',\bm k)\!=\!\iint d{\bm p'}d{\bm p}\,
\tilde\rho({\bm p'},{\bm p})
\langle{\bm k'}{\bm p'}\!\mid T\!\mid{\bm k}{\bm p} \rangle_{_{\!\cal A}}$,
representing the convolution between the
off-shell mixed density and a two-body interaction.
This general structure is shared by various formulations of
the optical-model 
potential~\cite{Watson1953,Kerman1959,Arellano1989,Chinn1995,Hebborn2023}.
In Ref.~\cite{Arellano2007a} it was
demonstrated that this general expression embodies 
a well defined functional structure in terms of the radial coordinate $z$,
characterized by the local density $\rho(z)$, 
leading to the expression in Eq.~\eqref{folding}. 
The second element lies in the momentum-space structure of the potential,
displaying negligible angular dependence when expressed in 
terms of momenta ${\bm K}$ and ${\bm q}$, together
with the weak dependence of $\tilde V(K,q)$ on $K$. 
The emerging complex nonlocality of hydrogenic type,
in conjunction with the radial form factors constitute 
intrinsic and well identifiable components of the 
microscopic potential.
These robust microscopically-driven findings offer new
ways to investigate nuclear reactions beyond
Woods-Saxon and Perey-Buck phenomenological prescriptions.
This is particularly relevant for current efforts aimed for the study
and description of nuclear reactions for 
the rare-isotope era~\cite{Hebborn2023}.

\vspace{12pt}
\noindent
\textit{Acknowledgments:}
H.F.A. acknowledges partial support provided by the supercomputing 
infrastructure of the NLHPC (ECM-02): Powered@NLHPC.
He also thanks the hospitality of colleagues of CEA-DAM at
Bruy\`eres-le-Ch\^atel, where part of this work was done.



\begin{thebibliography}{10}

\bibitem{Hebborn2023}
C~Hebborn, F~M Nunes, G~Potel, W~H Dickhoff, J~W Holt, M~C Atkinson, R~B Baker,
  C~Barbieri, G~Blanchon, M~Burrows, R~Capote, P~Danielewicz, M~Dupuis,
  Ch~Elster, J~E Escher, L~Hlophe, A~Idini, H~Jayatissa, B~P Kay, K~Kravvaris,
  J~J Manfredi, A~Mercenne, B~Morillon, G~Perdikakis, C~D Pruitt, G~H Sargsyan,
  I~J Thompson, M~Vorabbi, and T~R Whitehead.
\newblock Optical potentials for the rare-isotope beam era.
\newblock {\em Journal of Physics G: Nuclear and Particle Physics},
  50(6):060501, apr 2023.

\bibitem{Woods1954}
Roger~D. Woods and David~S. Saxon.
\newblock Diffuse surface optical model for nucleon-nuclei scattering.
\newblock {\em Phys. Rev.}, 95:577--578, Jul 1954.

\bibitem{Perey1962}
F.~Perey and B.~Buck.
\newblock A non-local potential model for the scattering of neutrons by nuclei.
\newblock {\em Nuclear Physics}, 32:353--380, 1962.

\bibitem{Koning2003}
A.J. Koning and J.P. Delaroche.
\newblock {Local and global nucleon optical models from 1 keV to 200 MeV}.
\newblock {\em Nuclear Physics A}, 713(3):231 -- 310, 2003.

\bibitem{Morillon2007}
B.~Morillon and P.~Romain.
\newblock Bound single-particle states and scattering of nucleons on spherical
  nuclei with a global optical model.
\newblock {\em Phys. Rev. C}, 76:044601, Oct 2007.

\bibitem{Tian2015}
Yuan Tian, Dan-Yang Pang, and Zhong-Yu Ma.
\newblock Systematic nonlocal optical model potential for nucleons.
\newblock {\em International Journal of Modern Physics E}, 24(01):1550006,
  2015.

\bibitem{Hagen2012}
G.~Hagen and N.~Michel.
\newblock Elastic proton scattering of medium mass nuclei from coupled-cluster
  theory.
\newblock {\em Phys. Rev. C}, 86:021602, Aug 2012.

\bibitem{Rotureau2017}
J.~Rotureau, P.~Danielewicz, G.~Hagen, F.~M. Nunes, and T.~Papenbrock.
\newblock Optical potential from first principles.
\newblock {\em Phys. Rev. C}, 95:024315, Feb 2017.

\bibitem{Idini2019}
A.~Idini, C.~Barbieri, and P.~Navr\'atil.
\newblock Ab initio optical potentials and nucleon scattering on medium mass
  nuclei.
\newblock {\em Phys. Rev. Lett.}, 123:092501, Aug 2019.

\bibitem{Rotureau2020}
Jimmy Rotureau.
\newblock Coupled-cluster computations of optical potential for medium-mass
  nuclei.
\newblock {\em Frontiers in Physics}, 8:285, 2020.

\bibitem{Arellano1995}
H.~F. Arellano, F.~A. Brieva, and W.~G. Love.
\newblock In-medium full-folding optical model for nucleon-nucleus elastic
  scattering.
\newblock {\em Phys. Rev. C}, 52:301--315, Jul 1995.

\bibitem{Amos2000}
K.~Amos, P.~J. Dortmans, H.~V. von Geramb, S.~Karataglidis, and J.~Raynal.
\newblock {\em {Advances in Nuclear Physics}}, volume~25 of {\em Advances in
  Nuclear Physics}.
\newblock Springer, New York, 2000.

\bibitem{Arellano2022}
H.~F. Arellano and G.~Blanchon.
\newblock {On the separability of microscopic optical model potentials and
  emerging bell-shape Perey-Buck nonlocality}.
\newblock {\em Eur. Phys. Journal A}, 59(1):1, 2022.

\bibitem{Wiringa1995}
R.~B. Wiringa, V.~G.~J. Stoks, and R.~Schiavilla.
\newblock {Accurate nucleon-nucleon potential with charge-independence
  breaking}.
\newblock {\em Phys. Rev. C}, 51(1):38--51, Jan 1995.

\bibitem{Arellano2002}
H.~F. Arellano and H.~V. von Geramb.
\newblock {Extension of the full-folding optical model for nucleon-nucleus
  scattering with applications up to 1.5 GeV}.
\newblock {\em Phys. Rev. C}, 66:024602, Aug 2002.

\bibitem{Ray1992}
L.~Ray, G.~W. Hoffmann, and W.~R. Coker.
\newblock Nonrelativistic and relativistic descriptions of proton-nucleus
  scattering.
\newblock {\em Physics Reports}, 212:223, 1992.

\bibitem{Aguayo2008}
F.~J. Aguayo and H.~F. Arellano.
\newblock Surface-peaked medium effects in the interaction of nucleons with
  finite nuclei.
\newblock {\em Phys. Rev. C}, 78:014608, 2008.

\bibitem{Arellano2015}
H.~F. Arellano and J.-P. Delaroche.
\newblock {Low-density homogeneous symmetric nuclear matter: Disclosing
  dinucleons in coexisting phases}.
\newblock {\em Eur. Phys. Journal A}, 51(1):7, January 2015.

\bibitem{Arellano1990b}
H.~F. Arellano, F.~A. Brieva, and W.~G. Love.
\newblock Role of nuclear densities in nucleon elastic scattering.
\newblock {\em Phys. Rev. C}, 42:652--658, Aug 1990.

\bibitem{Negele1970}
J.~W. Negele.
\newblock Structure of finite nuclei in the local-density approximation.
\newblock {\em Phys. Rev. C}, 1:1260--1321, Apr 1970.

\bibitem{Arellano2021}
H.~F. Arellano and G.~Blanchon.
\newblock {SWANLOP: Scattering waves off nonlocal optical potentials in the
  presence of Coulomb interactions}.
\newblock {\em Computer Physics Communications}, 259:107543, 2021.

\bibitem{Arellano2019}
H.~F. Arellano and G.~Blanchon.
\newblock {Exact scattering waves off nonlocal potentials under Coulomb
  interaction within Schr\"dinger's integro-differential equation}.
\newblock {\em Physics Letters B}, 789:256 -- 261, 2019.

\bibitem{Blumberg1966}
{Blumberg, L. N. and Gross, E. E. and VAN DER Woude, A. and Zucker, A. and
  Bassel, R. H.}
\newblock {Polarizations and Differential Cross Sections for the Elastic
  Scattering of 40-MeV Protons from $^{12}\mathrm{C}$, $^{40}\mathrm{Ca}$,
  $^{58}\mathrm{Ni}$, $^{90}\mathrm{Zr}$, and $^{208}\mathrm{Pb}$}.
\newblock {\em Phys. Rev.}, 147:812--825, Jul 1966.

\bibitem{Nadasen1981}
A.~Nadasen, P.~Schwandt, P.~P. Singh, W.~W. Jacobs, A.~D. Bacher, P.~T.
  Debevec, M.~D. Kaitchuck, and J.~T. Meek.
\newblock {Elastic scattering of 80-180 MeV protons and the proton-nucleus
  optical potential}.
\newblock {\em Phys. Rev. C}, 23:1023--1043, Mar 1981.

\bibitem{Seifert1993}
H.~Seifert, J.~J. Kelly, A.~E. Feldman, B.~S. Flanders, M.~A. Khandaker,
  Q.~Chen, A.~D. Bacher, G.~P.~A. Berg, E.~J. Stephenson, P.~Karen, B.~E.
  Norum, P.~Welch, and A.~Scott.
\newblock {Effective interaction for $^{16}\mathrm{O}$(p,p') and
  $^{40}\mathrm{Ca}$(p,p') at ${\mathit{E}}_{\mathit{p}}$=200 MeV}.
\newblock {\em Phys. Rev. C}, 47:1615--1635, Apr 1993.

\bibitem{Hutcheon1988}
D.A. Hutcheon, W.C. Olsen, H.S. Sherif, R.~Dymarz, J.M. Cameron, J.~Johansson,
  P.~Kitching, P.R. Liljestrand, W.J. McDonald, C.A. Miller, G.C. Neilson, D.M.
  Sheppard, D.K. McDaniels, J.R. Tinsley, P.~Schwandt, L.W. Swenson, and C.E.
  Stronach.
\newblock {The elastic scattering of intermediate energy protons from $^{40}$Ca
  and $^{208}$Pb}.
\newblock {\em Nuclear Physics A}, 483(3):429--460, 1988.

\bibitem{Entem2003}
D.~R. Entem and R.~Machleidt.
\newblock Accurate charge-dependent nucleon-nucleon potential at fourth order
  of chiral perturbation theory.
\newblock {\em Phys. Rev. C}, 68:041001, Oct 2003.

\bibitem{Blanchon2024}
G.~Blanchon and H.~F. Arellano.
\newblock {Microscopic bell-shape nonlocality: the case of proton scattering
  off $^{40}$Ca at 200 MeV}.
\newblock {\em arXiv:2402.15009}, 2024.

\bibitem{Watson1953}
Kenneth~M. Watson.
\newblock Multiple scattering and the many-body problem---applications to
  photomeson production in complex nuclei.
\newblock {\em Phys. Rev.}, 89:575--587, Feb 1953.

\bibitem{Kerman1959}
A.K Kerman, H~McManus, and R.M Thaler.
\newblock The scattering of fast nucleons from nuclei.
\newblock {\em Annals of Physics}, 8(4):551--635, 1959.

\bibitem{Arellano1989}
H.~F. Arellano, F.~A. Brieva, and W.~G. Love.
\newblock Full-folding-model description of elastic scattering at intermediate
  energies.
\newblock {\em Phys. Rev. Lett.}, 63:605--608, Aug 1989.

\bibitem{Chinn1995}
C.~R. Chinn, Ch. Elster, R.~M. Thaler, and S.~P. Weppner.
\newblock Propagator modifications in elastic nucleon-nucleus scattering within
  the spectator expansion.
\newblock {\em Phys. Rev. C}, 52:1992--2003, Oct 1995.

\bibitem{Arellano2007a}
H.~F. Arellano and E.~Bauge.
\newblock Functional medium dependence of the nonrelativistic optical model
  potential.
\newblock {\em Phys. Rev. C}, 76:014613, Jul 2007.

\end{thebibliography}

 \end{document}